\documentstyle[12pt]{article} 
\bibliography{plain}
\title{\bf Hierarchy of piecewise non-linear maps with non-ergodicity behavior}
\vspace{20mm}\author{M.A. Jafarizadeh$^{a,b,c}$\thanks{E-mail:
jafarizadeh@tabrizu.ac.ir}, M. Foroutan $^{b,c}$\thanks{E-mail:
m.foroutan@tabrizu.ac.ir} and S. Behnia $^{d,e}$\thanks{E-mail:
Sohrab-Behnia@hotmail.com}.
\\ $^a${\small Department of Theoretical Physics and Astrophysics,
Tabriz University, Tabriz 51664, Iran.}
\\ $^b${\small Institute for Studies in Theoretical Physics and
Mathematics, Teheran 19395-1795, Iran.}
\\ $^c${\small Excellency of Physics, Physics Department, Tabriz
University, Tabriz 51664, Iran.}
\\ $^d${\small Plasma Physics Research Center, IAU, Tehran 14835-159,
Iran.}
\\ $^e${\small Department of Physics, IAU, Ourmia, Iran.}}
\newpage
\begin{document} \maketitle \vspace{15mm}
\newpage
\begin{abstract}
We study the dynamics of hierarchy of piecewise maps generated by
one-parameter families of trigonometric chaotic maps and
one-parameter  families of elliptic chaotic maps of  $\mathbf{cn}$
and $\mathbf{sn}$ types, in detail. We calculate the Lyapunov
exponent and Kolmogorov-Sinai entropy of the these maps with
respect to control parameter. Non-ergodicity of these piecewise
maps is proven  analytically and investigated numerically . The
invariant measure of these maps which are not equal to one or
zero, appears to be characteristic of non-ergodicity behavior. A
quantity of interest is the Kolmogorov-Sinai entropy, where for
these maps are smaller than the sum of positive Lyapunov exponents
and
it confirms  the non-ergodicity of the maps.
 {\bf PACs
numbers:05.45.Ra, 05.45.Jn, 05.45.Tp }
\end{abstract}
\pagebreak \vspace{7cm}
\section{Introduction}
Ergodicity theory is a branch of dynamic systems dealing with
questions of average. Sometimes ergodic theory can make long-term
predictions about the average behavior  from initial data with
limited accuracy even for chaotic systems. The ergodic theory of
chaos has been studied in detail by Eckmann and Ruelle
\cite{Eck,Rue} and others, while non-ergodic mathematical models
scarcely exist. It is shown that simplest one-dimensional dynamic
systems satisfying the indecomposability assumption (and even the
assumption of topological transitivity) may be non-ergodic, which
shows that restrictions of this type are not quite reasonable in
the context of general dynamical systems \cite{Bla}. Also,
ergodicity has not been proven for some systems and sometimes we
have to face the problem of non-ergodicity of many chemical and
physical systems especially in solid systems (for example see
\cite{cu,sc,bo,ad}). It is shown that, candidate solid state
systems with extremely slow dynamics often have impurities
remaining from their synthesis  so there is always a suspicion
that non-ergodicity comes from disorder \cite{ar}. Also, there are
specific examples of arbitrary small perturbation to ergodic
systems that are not ergodic \cite{ma,do,tu}.
 The study of origins of the non-ergodicity and slow dynamics  of
 polymer gels \cite{ng1} and effects of temperature and swelling on
 chain dynamics  in sol-gel transition \cite{ng2,wu} and also
 non-ergodicity transition in colloidal gelation \cite{be} are
 some examples of research activity concerning non-ergodicity.\\
 The aim of present paper is twofold: To introduce the piecewise map
with its invariant measure and to clarify its non-ergodic
behavior. We present the hierarchy of one and many-parameter
families  of elliptic chaotic maps of $\mathbf{cn}$ and
$\mathbf{sn}$ types which can generate non-ergodic behavior. In
our definition of a piecewise map,  we assume that it consist of
components which can describe fixed point and chaotic  behavior
accordance with various values of parameter of the map. In
particular, we argue that this map satisfy the non-ergodic
assumption. The paper is organized as follows: The next section of
the paper will note the competing definitions of   piecewise
non-linear maps with complete boundary condition associated with
the one parameter families  of chaotic maps and one-parameter
families of elliptic chaotic maps of $\mathbf{cn}$ and
$\mathbf{sn}$ types. In Section 3, we present the invariant
measure  of piecewise maps and in Section 4 the Kolmogorov-Sinai
(KS) entropy and Lyapunov exponent (LE) of the map is studied. In
Section 5, we review  the ergodic theory and in Section 6, we give
our results consist  the KS-entropy and LE of the map and we shall
explain how this piecewise  maps behave non-ergodic. In addition
there will be a concluding section and two appendices.
\section{Piecewise non-linear maps}
\subsection{One-parameter families of trigonometric chaotic maps}
We first review  the one-parameter chaotic maps which are used to
construct  the piecewise map. The one-parameter chaotic maps
\cite{Ja1} are defined  as the ratio of polynomials of degree
$\mathbf{N}$:
$$\phi_{N}^{(1)}(x,\alpha)=\frac{\alpha^{2}(1+(-1)^{N}\quad
_{2}F_{1}(-N,N,\frac{1}{2},x))}{(\alpha
^{2}+1)+(\alpha^{2}-1)(-1)^{N}\quad_{2}F_{1}(-N,N,\frac{1}{2},x)}$$$$=\frac{\alpha^{2}(T_{N}(\sqrt{x}))^{2}}{1+(\alpha^{2}-1)(T_{N}(\sqrt{x})^{2})
}$$
$$\phi_{N}^{(2)}(x,\alpha)=\frac{\alpha^{2}(1-(-1)^{N}\quad_{2}F_{1}(-N,N,\frac{1}{2},(1-x)))}
{(\alpha^{2}+1)-(\alpha^{2}-1)(-1)^{N}
\quad_{2}F_{1}(-N,N,\frac{1}{2},(1-x))}$$
$$=\frac{\alpha^{2}(U_{N}(\sqrt{(1-x)}))^{2}}{1+(\alpha^{2}-1)(U_{N}(\sqrt{(1-x)})^{2})}$$
where $\mathbf{N}$ is an integer greater than one. Also,
$$_{2}F_{1}(-N,N,\frac{1}{2},x)=(-1)^{N}\cos (2N \arccos
\sqrt{x})=(-1)^{N} T_{2N}(\sqrt{x})$$ is the hypergeometric
polynomials of degree $N$ and $T_{N}(U_{n}(x))$ are  Chebyshev
polynomials of type I (type II), respectively. The conjugate maps
of the one-parameter families of chaotic maps which are used in
derivation of their invariant measure and calculation of their
KS-entropy are defined as:
$$\tilde{\phi}_{N}^{(1)}(x,\alpha)=h\circ \phi_{N}^{(1)}(x,\alpha)\circ
h^{-1}=\frac{1}{\alpha ^{2}}\tan ^{2}(N \arctan \sqrt{x})$$
$$\tilde{\phi}_{N}^{(2)}(x,\alpha)=h\circ \phi_{N}^{(2)}(x,\alpha)\circ
h^{-1}=\frac{1}{\alpha ^{2}}\cot ^{2}(N \arctan
\frac{1}{\sqrt{x}})$$ Conjugacy means that invertible map
$h(x)=\frac{1-x}{x}$ maps $I=[0,1]$ into $[0,\infty)$. To define
the piecewise maps constructed from one-parameter chaotic map, we
need to take into account boundary condition, namely we have to
choose the states on the phase space. Here we present examples of
these types which have been considered in the present
paper:$$\phi_{2}^{(1)}=\frac{\alpha^{2}(2x-1)^{2}}{4x(1-x)+\alpha^{2}(2x-1)^{2}}$$
$$\phi_{2}^{(2)}=\frac{4\alpha^{2}x(1-x)}
{1+4(\alpha^{2}-1)x(1-x)}$$
$$\phi_{3}^{(1)}=\phi_{3}^{(2)}=\frac{\alpha^{2}x(4x-3)^{2}}
{\alpha^{2}x(4x-3)^{2}+(1-x)(4x-1)^{2}}$$ Now, We introduce
hierarchy of piecewise maps generated by one-parameter families of
trigonometric chaotic maps. For
piecewise map $\phi_{N}^{(1)}$ with even $\mathbf{N}$,  we have:\\\\
$\phi_{N}^{(1)}(x,\alpha)=\left\{\begin{array}{ll}
\phi_{N}^{(1)}(x,\alpha_{1})&\alpha_{1}\in[N,\infty)\\
\phi_{N}^{(1)}(x,\alpha_{2})&\alpha_{2}\in[0,N]
\end{array}
\right.$\\
\\The range of the parameters $\alpha_{1}$ and $\alpha_{2}$  in the
maps $\phi_{N}^{(1)}(x,\alpha_{1})$ and
$\phi_{N}^{(1)}(x,\alpha_{2})$ are chosen to grantee,
respectively, chaotic behavior and two  fixed points at $x=0$ and
$x=1$. Figure 1 shows the plot of $\phi_{2}^{(1)}(x,\alpha)$ for
$\alpha_{1}=3$ and $\alpha_{2}=1$. In
$\phi_{N}^{(1)}(x,\alpha_{1})$ and $\phi_{N}^{(1)}(x,\alpha_{2})$,
$\mathbf{x}$ is limited respectively to:
$\acute{x}\in[0.152,0.848]$ and
$\ddot{x}\in[0,0.352]\cup[0.647,1]$. For given $y_{0}=0.5$ the
$\acute{x}$ are the roots of $\phi_{N}^{(1)}(x,\alpha_{1})= y_{0}$
and similarly  the $\ddot{x}$ are the roots of
$\phi_{N}^{(1)}(x,\alpha_{2})= y_{0}$.
\\For piecewise map
$\phi_{N}^{(2)}(x,\alpha)$ with even $\mathbf{N}$,  we have:\\\\
$\phi_{N}^{(2)}(x,\alpha)=\left\{\begin{array}{ll}
\phi_{N}^{(2)}(x,\alpha_{1})&\alpha_{1}\in[0,\frac{1}{N}]\\
\phi_{N}^{(2)}(x,\alpha_{2})&\alpha_{2}\in[\frac{1}{N},\infty)
\end{array}\right.$\\
\\The range of the parameters $\alpha_{1}$ and $\alpha_{2}$  in the
maps $\phi_{N}^{(2)}(x,\alpha_{1})$ and
$\phi_{N}^{(2)}(x,\alpha_{2})$ are chosen to grantee,
respectively, two fixed points at $x=0$ and $x=1$ and chaotic
behavior. Figure 2 shows the plot of $\phi_{2}^{(2)}(x,\alpha)$
for $\alpha_{1}=0.25$ and $\alpha_{2}=0.75$. In
$\phi_{N}^{(2)}(x,\alpha_{1})$ and $\phi_{N}^{(2)}(x,\alpha_{2})$,
$\mathbf{x}$ is limited respectively to:
$\acute{x}\in[0,0.378]\cup[0.621,1]$ and $\ddot{x}\in[0.2,0.8]$.
For given $y_{0}=0$ the $\acute{x}$ are the roots of
$\phi_{N}^{(2)}(x,\alpha_{1})= y_{0}$ and similarly the $\ddot{x}$
are the roots of $\phi_{N}^{(2)}(x,\alpha_{2})= y_{0}$. \\For
piecewise map
$\phi_{N}^{(1,2)}$ with odd $\mathbf{N}$,  we have:\\\\
$\phi_{N}^{(1,2)}(x,\alpha)=\left\{\begin{array}{ll}
\phi_{N}^{(1,2)}(x,\alpha_{1})&\quad\quad\alpha_{1}\in[\frac{1}{N},N]\\
\phi_{N}^{(1,2)}(x,\alpha_{2})&\quad\quad\alpha_{2}\in[0,\frac{1}{N}]\cup[N,\infty)
\end{array}\right.$\\
\\This map has chaotic behavior for $\alpha_{2}$ and has a fixed
point in $x=0$  for $\alpha_{1}$. Figure 3 shows the plot of
$\phi_{3}^{(1,2)}(x,\alpha)$ for $\alpha_{1}=1.5$ and
$\alpha_{2}=0.2$. In $\phi_{N}^{(1,2)}(x,\alpha_{1})$ and
$\phi_{N}^{(1,2)}(x,\alpha_{2})$, $\mathbf{x}$ is limited
respectively to: $\acute{x}\in[0,0.2]\cup[0.315,1]$ and
$\ddot{x}\in[0.086,0.453]\cup[0.947,1]$. For given $y_{0}=0.5$ the
$\acute{x}_{i}$ are the roots of $\phi_{N}^{(2)}(x,\alpha_{1})=
y_{0}$ and similarly the $\ddot{x}_{i}$ are the roots of
$\phi_{N}^{(2)}(x,\alpha_{2})= y_{0}$.
\subsection{ One-parameter families of elliptic chaotic maps
 of cn and sn types}
Here we first review a hierarchy of one-parameter families of
elliptic  of $\mathbf{cn}$  and $\mathbf{sn}$ types that have been
used for constructing the piecewise maps with non-ergodic
behavior. These kinds of maps are defined as the ratios of
Jacobian elliptic functions of $\mathbf{cn}$ and $\mathbf{sn}$
types through the following equation \cite{Ja2}:
 $$\phi_{N}^{(1)}(x,\alpha)=\frac{\alpha^2(cn(N cn^{-1}(\sqrt{x})))^{2}}{1+(\alpha^{2}-1)(cn(N cn^{-1}(\sqrt{x})))^{2}}$$
$$\phi_{N}^{(2)}(x,\alpha)=\frac{\alpha^2(sn(N sn^{-1}(\sqrt{x})))^{2}}{1+(\alpha^{2}-1)(sn(N sn^{-1}(\sqrt{x})))^{2}}$$
where $\alpha$ is control parameter. For $N=2$ we have:
$$\phi_{2}^{(1)}(x,\alpha)=\frac{\alpha^2((1-k^{2})(2x-1)+k^{2}x^{2})^2}{(1-k^{2}+2k^{2}x-k^{2}x^{2})^{2}+(\alpha^{2}-1)((1-k^{2})(2x-1)+k^{2}x^{2})^{2}}$$
$$\phi_{2}^{(2)}(x,\alpha)=\frac{4\alpha^2x(1-k^{2}x)(1-x)}{(1-k^{2}x^{2})^{2}+4x(1-x)(\alpha^{2}-1)(1-k^{2}x)}$$
It has been proved \cite{Ja2} that for small values of the
parameter $\mathbf{K}$ of the elliptic function, these maps are
topologically conjugate to the one parameter families of chaotic
maps. Similar to the introduced piecewise in previous section
(2.1), the piecewise of elliptic maps can be introduced. For an
example for piecewise elliptic map $\phi_{2}^{(2)}(x,\alpha)$, we have:\\\\
$\phi_{2}^{(2)}(x,\alpha)=\left\{\begin{array}{ll}
\phi_{2}^{(2)}(x,\alpha_{1})&\alpha_{1}\in[0,\frac{1}{N}]\\
\phi_{2}^{(2)}(x,\alpha_{2})&\alpha_{2}\in[\frac{1}{N},\infty)
\end{array}\right.$\\
\\The range of the parameters $\alpha_{1}$ and $\alpha_{2}$  in the
maps $\phi_{2}^{(2)}(x,\alpha_{1})$ and
$\phi_{2}^{(2)}(x,\alpha_{2})$ are chosen to grantee,
respectively, two  fixed points at $x=0$ and $x=1$ and chaotic
behavior. Figure 4 shows the plot of the elliptic map
$\phi_{2}^{(2)}(x,\alpha)$ for $\alpha_{1}=0.5$ and
$\alpha_{2}=2.5$. In $\phi_{2}^{(2)}(x,\alpha_{1})$ and
$\phi_{2}^{(2)}(x,\alpha_{2})$, $\mathbf{x}$ is limited
respectively to: $\acute{x}\in[0,0.28]\cup[0.72,1]$ and
$\ddot{x}\in[0.027,0.973]$. For given $y_{0}=0.5$ the $\acute{x}$
are the roots of $\phi_{2}^{(2)}(x,\alpha_{1})= y_{0}$ and
similarly the $\ddot{x}_{i}$ are the roots of
$\phi_{2}^{(2)}(x,\alpha_{2})= y_{0}$.

\section{Invariant measure}
Invariant measure or SRB measure is supported on  an attractor and
describes the statistical of long-time behavior of the orbits with
respect to Lebesgue measure. For invariant measure of
$\phi_{N}^{(i)}$ map ($i=1,2$) satisfying the Frobenius-Perron
(FP) operator \cite{aba}, we have:
$$\mu(y)=\int_{0}^{1}\delta(y-\phi_{N}^{(i)}(x,\alpha))\mu(x) dx$$
which is equivalent to:
\begin{equation}\mu(y)=\sum_{x\in\phi_{N}^{-1(i)}(y,
\alpha)}\mu(x)\frac{dx}{dy}\end{equation} For chaotic part of the
piecewise map, i.e.; $y\in[0,y_{0}]$ for
$\phi_{2}^{(1)}(x,\alpha)$ and $y\in[y_{0},1]$ for both
$\phi_{3}^{(1,2)}(x,\alpha)$ and $\phi_{2}^{(2)}(x,\alpha)$, the
invariant measure $\mu(x,\beta)$ is defined as:
\begin{equation}\frac{1}{\pi}\frac{\sqrt{\beta}}{\sqrt{x(1-x)}(\beta+(1-\beta)x)}\end{equation}
With $\beta>0$ is the invariant measure of the maps
$\phi_{N}^{(i)}(x,\alpha)$ provided that, we choose the parameter
$\alpha$ in the following form:
\begin{equation}\alpha=\frac{\sum_{k=0}^{[\frac{N-1}{2}]}C_{2k+1}^{N}\beta^{-k}}{\sum_{k=0}^{[\frac{N}{2}]}C_{2k}^{N}\beta^{-k}}\end{equation}
in $\phi_{N}^{(i)}(x,\alpha)$ maps for odd values of $\mathbf{N}$,
and; \
\begin{equation}\alpha=\frac{\beta\sum_{k=0}^{[\frac{N}{2}]}C_{2k}^{N}\beta^{-k}}{\sum_{k=0}^{[\frac{N-1}{2}]}C_{2k+1}^{N}\beta^{-k}}\end{equation}
in $\phi_{N}^{(i)}(x,\alpha)$ maps for even values of
$\mathbf{N}$, where the symbol $[\quad]$ means greatest integer
part. \\
As it is shown in Appendix A, for satisfying the invariant measure
$\mu(x,\beta)$ with $\alpha$, we obtain:
\begin{equation}\alpha=\frac{B(\frac{1}{\beta})}{A(\frac{1}{\beta})}\end{equation}
with polynomials $A(x)$ and $B(x)$ defined as:
\begin{equation}A(x)=\sum_{k=0}^{[\frac{N}{2}]}C_{2k}^{N}x^{k}\end{equation}\begin{equation}B(x)=\sum_{k=0}^{[\frac{N-1}{2}]}C_{2k+1}^{N}x^{k}\end{equation}
If an invariant measure can be indecomposed into parts that are
invariant, the measure is called non-ergodic. There may be several
invariant measure for a dynamical system. If there is a fixed
point $x^{\ast}$, then a point distribution $\delta(x-x^{\ast})$
in that point is an invariant measure, even if the fixed point is
unstable. Therefore for the fixed point part of the piecewise map
i.e., $y\in[y_{0},1]$ for $\phi_{2}^{(1)}(x,\alpha)$ and
$y\in[0,y_{0}]$ for both maps $\phi_{3}^{(1,2)}(x,\alpha)$ and
$\phi_{2}^{(2)}(x,\alpha)$,  the average density measure
$\mu(x,\beta)$ has the following asymptotic form of the delta
function as $\alpha$ goes to the zero and one, respectively;
$$\mu_{av}(x,\alpha)\stackrel{\rm \alpha\rightarrow
0}\longrightarrow\delta(x)$$
$$\mu_{av}(x,\alpha)\stackrel{\rm \alpha\rightarrow
1}\longrightarrow\delta(x-1)$$ where the first one corresponds to
invariant measure associated with the fixed point at $x=0$ and the
latter one corresponds to fixed point at $x=1$.\\Since for small
values of $\mathbf{K}$ the parameter of the elliptic function, the
elliptic chaotic maps of $\mathbf{cn}$ and $\mathbf{sn}$ types are
topologically conjugate to the one-parameter families of
trigonometric chaotic maps, we can obtain the invariant measure of
these maps for small $\mathbf{K}$ \cite{Ja4}. As $\mathbf{K}$
vanishes, these maps are to trigonometric chaotic maps.
\section{Kolmogorov-Sinai Entropy and Lyapunov exponents}
Kolmogorov-Sinai (KS) entropy and Lyapunov characteristic
exponents are two related ways of measuring 'disorder' in a
dynamic system. The definition of them can be found in many
textbooks \cite{dor}. To calculate KS-entropy here, we use the
fact that it  is equal to: $$h(\mu,\phi_{N}^{(i)}(x,\alpha)=\int
\mu(x) dx \ln|\frac{d}{dx}\phi_{N}^{(i)}(x,\alpha)|$$ which is
also a statistical mechanical expression for the Lyapunov
characteristic, that is mean divergence rate of two nearby orbits.
As it is shown in Appendix B, the KS-entropy of
$\phi_{N}^{(i)}(x,\alpha)$ has following expression:
$$h(\mu,\phi_{N}^{(i)}(x,\alpha))=\ln(\frac{N(1+\beta+2\sqrt{\beta})^{N-1}}
{(\sum_{k=0}^{[\frac{N}{2}]}C_{2k}^{N}\beta^{k})(\sum_{k=0}^{[\frac{N-1}{2}]}C_{2k+1}^{N}\beta^{k})})$$
A useful numerical way to characterize chaotic phenomena in
dynamic systems is by means of the Lyapunov exponents that
describe the separation rate of systems whose initial conditions
differ by a small perturbation. Suppose that there is a small
change $\delta x(0)$ in the initial state $x(0)$. At time
$\mathbf{t}$ this has changed to $\delta x(t)$ given by:
\begin{equation}
\delta x(t)\approx\delta x(0)|\frac{d\phi'}{dx}(x(0))|=\delta
x(0)|\phi'(x(t-1)).\phi'(x(t-2)).......\phi'(x(0))|,
\end{equation}
Where we have used the chain rule to expand the derivative of
$\mathbf{\phi}$. In the limit of infinitesimal perturbations
$\delta x(0)$ and infinite time we get an average exponential
amplification, the Lyapunov exponent $\lambda$ ,
\begin{equation}
\lambda=lim_{t\rightarrow\infty}\frac{1}{t}\ln|\frac{\delta
x(t)}{\delta
x(0)}|=lim_{t\rightarrow\infty}\frac{1}{t}\ln|\frac{d\phi'}{dx}(x(0))|=lim_{t\rightarrow\infty}\sum_{k=0}^{t-1}\ln|\phi'(x(k))|
\end{equation}
\section{Ergodicity and non-ergodicity}
An probabilistic dynamic system is characterized as ergodic or
non-ergodic by its marginal probability distributions. If the
distributions have e.g. infinite variances so that a process mean
cannot be defined, then the system is non-ergodic. An ergodic
system has "convergent" qualities over time, variances are finite
and a non-time-dependent process mean is clearly defined. Here a
brief description of ergodic theory of chaos \cite{Fa} is
presented: let ($\Omega$, $F$, $\mu$) be a probability space,
$\Omega$ is the sample space, i.e.; the space of points, $\omega$
designating the elementary outcomes of an experiment. F is the
$\sigma$-field (or $\sigma$-algebra) of events. An event is a set
A$\subset$$\Omega$ which is of interest. the $\sigma$-field F is
the ensemble of all events, i.e., A$\in$F. Also $\mu$ designates a
probability measure of F. A transformation T is ergodic, if it has
the probability that for almost every $\omega$, the orbit
$\{\omega, T\omega, T^{2}\omega, ...\}$ of $\omega$ is a sort of
replica of $\Omega$ itself. Formally, we shall say that T is
ergodic if each invariant set A, i.e.; a set such that
T$^{-1}$(A)=A, is trivial in the sense that it has measure either
zero or one. T$^{-1}$(A)=A $\Rightarrow$$\mu$(A)=0 or $\mu$(A)=1.
The transformation T is ergodic (or indecomposable, or metrically
transitive), if in the Birkhoff theorem, for any integrable,
real-value function f, the limit value $\hat{f}$ is constant and
we have $\mu$-almost everywhere.
$$\lim_{n\rightarrow\infty}\frac{1}{n}\sum_{k=0}^{n-1}f(T^{k}\omega)d\mu(\omega)
=\int_{\Omega}f(\omega)d\mu(\omega)$$ In this case, the average
value of $f(.)$, evaluated along the orbit $T^{k}\omega$,
converges $\mu$-almost everywhere to the mathematical expecting or
mean of $f(.)$, evaluated on the space $\Omega$. In other word,
for ergodic systems, the time average is equal to the space (or
phase) average. One further consideration should be added at this
point. The equality of KS-entropy and sum of all positive Lyapunov
exponents;
$$h_{KS}=\sum_{\lambda_{l}>0}\lambda_{l}$$
indicates that in chaotic region, this map is ergodic as Birkhof
ergodic theorem predicts \cite{kel}. In other words, when the
KS-entropy is smaller than the sum of positive LE, the  map  has
characterization of a non-ergodic behavior.
\\Also, studied based on invariant measure analysis can be
useful for confirming the non-ergodicity behavior of a map. For a
non-ergodic system we have:
$$\mu^{-1}(A)=\{x\in[0,1]\quad y=M(x)\quad y\in A\}$$
$$\mu(A)<1$$
i.e., the invariant measure which is not equal to zero or one,
appears to be characteristic of non-ergodic behavior.
\section{Results and discussion}
In this section we present the results of the analysis numerically
for piecewise map. Figure 5, 6, 7 and 8 show the variation of LE
and KS-entropy with the parameter $\alpha$. A positive LE implies
that two nearby trajectories exponentially diverge (at last
locally). Negative LE indicate contraction along certain
directions, and zero LE indicate that along the relevant
directions there is neither expansion nor contraction.
\\In figure 5, the LE and the corresponding KS-entropy's have been
shown for $\phi_{2}^{(1)}(x,\alpha_{1})$ and
$\phi_{2}^{(1)}(x,\alpha_{2})$ by some points. A quantity  of
interest is that the KS-entropy is smaller than the sum of
positive LE for piecewise maps. Because of this relation, it is
clear that this map  has characterization of a non-ergodic
behavior.\\The invariant measure of this map is equal to 0.696
which is smaller than one, therefore this map behaves non-ergodic.
The above analysis are presented for $\phi_{2}^{(2)}(x,\alpha)$,
$\phi_{3}^{(1,2)}(x,\alpha)$ and the elliptic map
$\phi_{2}^{(2)}(x,\alpha)$ (see figures 5-8 ).The invariant
measure of these maps, respectively  are equal to 0.6, 0.314,
0.946 which confirm the non-ergodicity behavior of introduced
piecewise maps.

\section{Conclusion}
A recent attempts in introducing the hierarchy of chaotic maps
 with their invariant measure \cite{Ja1,Ja2,Ja3,Ja4,Ja5} allows us to advance in answering to a question
how to define non-ergodic maps and what are the condition for
non-ergodicity in these types of system.\\ In this paper we
introduce the piecewise maps with their invariant measure. Our
numerically calculation shows that values of the KS-entropy are
smaller than the sum of positive LE for piecewise maps, therefore
these maps behaves non-ergodic.  Following, the non-ergodicity
behavior, also can be confirmed by the invariant measure which is
not equal to zero or one.
\section{Appendix A}\setcounter{equation}{0}
Similar to the calculation of the invariant measure in our
pervious papers \cite{Ja1,Ja2,Ja3,Ja4,Ja5}, we present here it for
the piecewise chaotic map.
 In order to prove that measure (3-2) satisfied equation (3-1), we
consider the conjugate map;
\begin{equation}\tilde{\phi}_{N}^{(1)}(x,\alpha)=\frac{1}{\alpha ^{2}}\tan
^{2}(N \arctan \sqrt{x})\end{equation} with measure
$\tilde{\mu}_{\tilde{\phi}_{N}}$ related to the measure
$\mu_{\phi_{N}}$ with the following relation:
$$\tilde{\mu}_{\tilde{\phi}_{N}}(x)=\frac{1}{(1+x)^{2}}\mu_{\phi_{N}}(\frac{1}{1+x}).$$
Denoting $\tilde{\phi}_{N}(x,\alpha)$ on the left-hand side of
(8-1) by  $\mathbf{y}$ and inverting it, we get:
\begin{equation}x_{k}=\tan^{2}(\frac{1}{N}\arctan
\sqrt{y\alpha^{2}}+\frac{k\pi}{N}), \quad k=1,...N.\end{equation}
Then, taking derivative of $x_{k}$ with respect to $\mathbf{y}$,
we obtain
\begin{equation}
|\frac{dx_{k}}{dy}|=\frac{\alpha}{N}\sqrt{x_{k}}(1+x_{k})\frac{1}{\sqrt{y}(1+\alpha^{2}y)}.
\end{equation}
Substituting the above result in equation (3-1), we get:
\begin{equation}
\tilde{\mu}_{\tilde{\phi}_{N}}(y)\sqrt{y}(1+\alpha^{2}y)=\frac{\alpha}{N}\sum_{k}\sqrt{x_{k}}(1+x_{k})\tilde{\mu}_{\tilde{\phi}_{N}}(x_{k}).
\end{equation}
Now, by considering the following ansatz for the invariant measure
$\tilde{\mu}_{\tilde{\phi}_{N}}(y)$:
\begin{equation}
\tilde{\mu}_{\tilde{\phi}_{N}}(y)=\frac{\sqrt{\beta}}{\sqrt{y}(1+\beta
y)},\end{equation}the above equation reduces to:
$$\frac{1+\alpha^{2}y}{1+\beta y}=\frac{\alpha}{N}\sum_{k=1}^{N}(\frac{1+x_{k}}{1+\beta x_{k}})$$
which can be written as:
\begin{equation}
\frac{1+\alpha^{2}y}{1+\beta
y}=\frac{\alpha}{\beta}+(\frac{\beta-1}{\beta^{2}})\frac{\partial}{\partial\beta^{-1}}\left(\begin{array}{c}\ln(\prod_{k=1}^{N}(\beta^{-1}+x_{k}))\end{array}\right)
.
\end{equation}
To evaluate the second term in the right-hand side of above
formulas, we can write the equation in the following form:
$$0=\alpha^{2} y \cos^{2}(N\arctan\sqrt{x})-\sin^{2}(N\arctan\sqrt{x})$$
$$=\frac{(-1)^{N}}{(1+x)^{N}}\left(\begin{array}{c}\alpha^{2}y(\sum_{k=0}^{[\frac{N}{2}]}C_{2k}^{N}(-1)^{N}x^{k})^{2}-x(\sum_{k=0}^{[\frac{N-1}{2}]}C_{2k+1}^{N}(-1)^{N}x^{k})^{2}\end{array}\right)
=\frac{constant}{(1+x)^{N}}\prod_{k=1}^{N}(x-x_{k}),$$ where
$x_{k}$ are the roots of equation (8-1) and they are given by
formula (8-2). Therefore, we have:
$$\frac{\partial}{\partial\beta^{-1}}\ln\left(\begin{array}{c} \prod_{k=1}^{N}(\beta^{-1}+x_{k})
\end{array}\right)
=$$
$$\frac{\partial}{\partial\beta^{-1}}\ln\left(\begin{array}{c}(1-\beta^{-1})^{N}(\alpha^{2}y
\cos^{2}(N\arctan\sqrt{-\beta^{-1}})-\sin^{2}(N\arctan\sqrt{-\beta^{-1}})\end{array}\right)=$$
\begin{equation}
-\frac{N\beta}{\beta-1}+\frac{\beta N
(1+\alpha^{2}y)A(1/\beta)}{(A(1/\beta))^{2}\beta^{2}y+(B(1/\beta))^{2}}.
\end{equation}
In deriving the above formula, we have used the following
identities:
\begin{equation}
\cos(N\arctan \sqrt{x})=\frac{A(-x)}{(1+x)^{N/2}},\quad\quad\quad
\sin(N\arctan\sqrt{x})=\sqrt{x}\frac{B(-x)}{(1+x)^{N/2}}.
\end{equation}
inserting the result (8-7) in (8-6), we get:
$$\frac{1+\alpha^{2}y}{1+\beta y}=\frac{1+\alpha^{2}y}{B(1/\beta)/A(1/\beta)+\beta(\alpha A(1/\beta)/B(1/\beta))y}$$
Hence, to get the final result, we have to choose the parameter
$\alpha$ as:  $$\alpha=\frac{B(1/\beta)}{A(1/\beta)}.$$
\section{Appendix B}\setcounter{equation}{0}
The KS entropy of one-parameter families of chaotic map is given
by:
$$h(\mu,\phi(x,\alpha))=\int\mu(x)dx\ln|\frac{d}{dx}\phi(x,\alpha)|$$where
$$\varphi(x,\alpha)=y=\frac{1}{\alpha^{2}} (tan^{2}(N \arctan \sqrt{x}))$$
Therefore to calculate $h(\mu,\varphi(x,\alpha))$ we have:

$$h(\mu,\varphi(x,\alpha))=\int_{0}^{\infty}\tilde{\mu}(x) dx \ln
\left(\left|\frac{N}{\alpha^{2}}\frac{1}{\sqrt{x}(1+x)}\frac{\sin
N(\arctan \sqrt{x})}{\cos^{3}N(\arctan \sqrt{x})}\right|\right)$$
using the relation (8-8), we get:
\begin{equation}h(\mu,\varphi(x,\alpha))=\frac{1}{\pi}\int_{0}^{\infty}\frac
{\sqrt\beta dx}{\sqrt{x}(1+\beta x)}\ln
\left(\left|\frac{N}{\alpha^{2}}\frac{(1+x)^{N-1}B(-x)}{(A(-x))^3}\right|\right)\end{equation}
We see that polynomials appearing in the numerator (denominator)
of the integrand appearing on the right-hand side of equation
(9-1), have $\frac{1}{2}[N-1](\frac{1}{2}[N])$ simple roots,
denoted by $x_{k}^{B}$, $k=1,...,\frac{1}{2}[N-1]$($x_{k}^{A}$,
$k=1,...,\frac{1}{2}[N]$) in the interval $[0,\infty)$. Hence we
can writ the above formula in the following form:
$$h(\mu,\varphi(x,\alpha))=\frac{1}{\pi}\int_{0}^{\infty}\frac
{\sqrt\beta dx}{\sqrt{x}(1+\beta x)}\ln
\left(\frac{N}{\alpha^{2}}\frac{(1+x)^{N-1}\prod_{k=1}^{[(N-1)/2]}|x-x_{k}^{B}|}{\prod_{k=1}^{[N/2]}|x-x_{k}^{A}|}
\right)$$ Now making the following  change of variable$
\sqrt{\beta}x=\tan\theta$, and taking into account that degree of
numerators and denominator are equal for both even and odd values
on $\mathbf{N}$we get:
$$h(\mu,\varphi(x,\alpha))=\frac{1}{\pi}\int_{0}^{\infty}d\theta\quad\{\ln(\frac{N}{\alpha^{2}})
+(N-1)\ln|\beta+1+(\beta-1)\cos\theta|$$$$+\Sigma_{k=1}^{[(N-1)/2]}\ln|1-x_{k}^{B}\beta+(1+x_{k}^{B}\beta)\cos\theta|
-3\Sigma_{k=1}^{[N/2]}\ln|1-x_{k}^{A}\beta+(1+x_{k}^{A}\beta)\cos\theta|\}$$
Using the following integrals:$$
\frac{1}{\pi}\int_{0}^{\pi}\ln|a+b\cos\theta|=
\left\{\begin{array}{llll} \ln|\frac{a+\sqrt{a^{2}-b^{2}}}{2}|  & & &  |a|>|b| \\
\ln|\frac{b}{2}|  &  &  & |a|\leq|b| \end{array} \right.
$$

\newpage
Figures captions\\
\\FIG. 1. plot of $\phi_{2}^{(1)}(x,\alpha)$ for $\alpha_{1}=3$
 and $\alpha_{2}=1$.
\\FIG. 2. plot of $\phi_{2}^{(2)}(x,\alpha)$ for $\alpha_{1}=0.25$
and $\alpha_{2}=0.75$.
\\FIG. 3. plot of $\phi_{3}^{(1,2)}(x,\alpha)$ for $\alpha_{1}=1.5$
and $\alpha_{2}=0.2$.
\\FIG. 4. plot of the elliptic map $\phi_{2}^{(2)}(x,\alpha)$ for $\alpha_{1}=0.5$
and $\alpha_{2}=2.5$.
\\FIG. 5. The variation of the Lyapunov exponent (dotted curve) and the
KS-entropy (solid curve) of the $\phi_{2}^{(1)}(x,\alpha)$ with
the  parameter $\alpha$. $\Box$ and $\star$  show the valuates of
Lyapunov exponent and KS-entropy, respectively for
$\phi_{2}^{(1)}(x,\alpha_{1})$ and $\phi_{2}^{(1)}(x,\alpha_{2})$.
\\FIG. 6. The variation of the Lyapunov exponent (dotted curve) and the
KS-entropy (solid curve) of the $\phi_{2}^{(2)}(x,\alpha)$ with
the parameter $\alpha$. $\Box$ and $\star$  show the valuates of
Lyapunov exponent and KS-entropy, respectively for
$\phi_{2}^{(2)}(x,\alpha_{1})$ and $\phi_{2}^{(2)}(x,\alpha_{2})$.
\\FIG. 7. The variation of the Lyapunov exponent (dotted curve) and the
KS-entropy (solid curve) of the $\phi_{3}^{(1,2)}(x,\alpha)$ with
the  parameter $\alpha$. $\Box$ and $\star$  show the valuates of
Lyapunov exponent and KS-entropy, respectively for
$\phi_{3}^{(1,2)}(x,\alpha_{1})$ and
$\phi_{3}^{(1,2)}(x,\alpha_{2})$.
\\FIG. 8. The variation of the Lyapunov exponent (solid curve) of the elliptic map
$\phi_{2}^{(2)}(x,\alpha)$ with the parameter $\alpha$ for $K=0$.
$\Box$ and $\star$  show the valuates of Lyapunov exponent and
KS-entropy, respectively for $\phi_{2}^{(2)}(x,\alpha_{1})$ and
$\phi_{2}^{(2)}(x,\alpha_{2})$.

\end{document}